\documentclass[twocolumn,pra,aps,showpacs,superscriptaddress,floatfix,showkeys,10pt]{revtex4-1}

\usepackage{wrapfig,lipsum,booktabs}
\usepackage{array}
\usepackage[colorlinks=true,linkcolor=blue,citecolor=magenta,urlcolor=blue]{hyperref}

\usepackage{amsmath,amsfonts}
\usepackage[]{graphics,graphicx,epsfig,wrapfig}
\usepackage{latexsym}
\usepackage{braket}
\usepackage{amssymb}
\usepackage{float}
\usepackage{graphicx}         
\usepackage{color}

\usepackage{amsmath,amsfonts,amssymb,amsthm}
\usepackage{color,graphicx}
\usepackage{bbm,bm}
\usepackage{booktabs}
\usepackage{multirow}
\usepackage{subcaption}
\usepackage{caption}

\usepackage{pifont}

\newtheorem{result}{Result}
\newtheorem{lemma}{Lemma}
\newtheorem{proposition}{Proposition}
\newtheorem{observation}{Observation}

\def\begeq{\begin{equation}}
\def\endeq{\end{equation}}
\def\begres{\begin{result}}
\def\endres{\end{result}}
\def\beglem{\begin{lemma}}
\def\endlem{\end{lemma}}
\def\begprop{\begin{proposition}}
\def\endprop{\end{proposition}}
\def\begobs{\begin{observation}}
\def\endobs{\end{observation}}

\begin{document}
	
	\title{Multiparty quantum random access codes}
	
	\author{Debashis Saha}
	\affiliation{Institute of Theoretical Physics and Astrophysics, National Quantum Information Center, Faculty of Mathematics, Physics and Informatics, 80-308, Gdansk, Poland}
	\affiliation{Center for Theoretical Physics, Polish Academy of Sciences, Al. Lotnik\'{o}w 32/46, 02-668 Warsaw, Poland}
	\author{Jakub J. Borka{\l}a} 
	\affiliation{Institute of Theoretical Physics and Astrophysics, National Quantum Information Center, Faculty of Mathematics, Physics and Informatics, 80-308, Gdansk, Poland}

	
\begin{abstract}

Random access code (RAC), a primitive for many information processing protocols, enables one party to encode $n$-bit string into one bit of message such that another party can retrieve partial information of that string. We introduce the multiparty version of RAC in which the $n$-bit string is distributed among many parties. For this task, we consider two distinct quantum communication scenarios: one allows shared quantum entanglement among the parties with classical communication, and the other allows communication through a quantum channel. We present several multiparty quantum RAC protocols that outclass its classical counterpart in both the aforementioned scenarios.

\end{abstract}

\maketitle
	
\section{Introduction}
	
The well known Holevo bound \cite{holevo,book} states that  the potential information carried by a $d$-dimensional quantum system is no more than $\log d$ bits. Despite this fact, quantum communication overshadows classical communication in several aspects. One of the prevailing manifestations of quantum communication advantage has been reported under the scope of \textit{random access codes} (RAC) \cite{rac1,rac2,rac3}. In the simplest form of RAC, the sender encodes $n$-bit string into a two-dimensional system such that the receiver can extract one randomly chosen bit out of the $n$ bits with as high probability as possible \cite{QRAC}. The resources employed in this task are: the communicated system (which are restricted to certain dimension) from sender to receiver, and the pre-shared randomness prior to the task. In the classical regime, both the communication channel and shared randomness are classical. While one may exploit quantum resources in two ways: (1) the sender encodes the inputs in a quantum system, which we refer to as \textit{Quantum random access codes} (QRAC); (2) the sender and receiver share quantum correlation (quantum entanglement) while the sender communicates via a classical system, which we refer to \textit{Entanglement assisted random access codes} (EARAC). Remarkably, these two ways of implementing quantum resources are not equivalent in general \cite{earac,magic7,SvsS,drac}. Besides the conventional applications of QRAC in quantum key-distribution \cite{qkd,qkd1,qkd2,sec,2s2r}, quantum randomness certification \cite{rc} and dimension witness of quantum systems \cite{racd,EM,BBP,brac,galvao}, the standard RAC task has been adapted to many novel communication problems yielding significant results \cite{ic,bobby,racbox,prac,opur,st1,st2,st3}. However, a generalization of RAC to the multiparty scenario remains unexplored. The aim of this article is to provide first instances of \textit{multiparty random access code} protocol under quantum and classical communication.

\begin{figure}[H]
 \centering
  \includegraphics[width=0.45\textwidth]{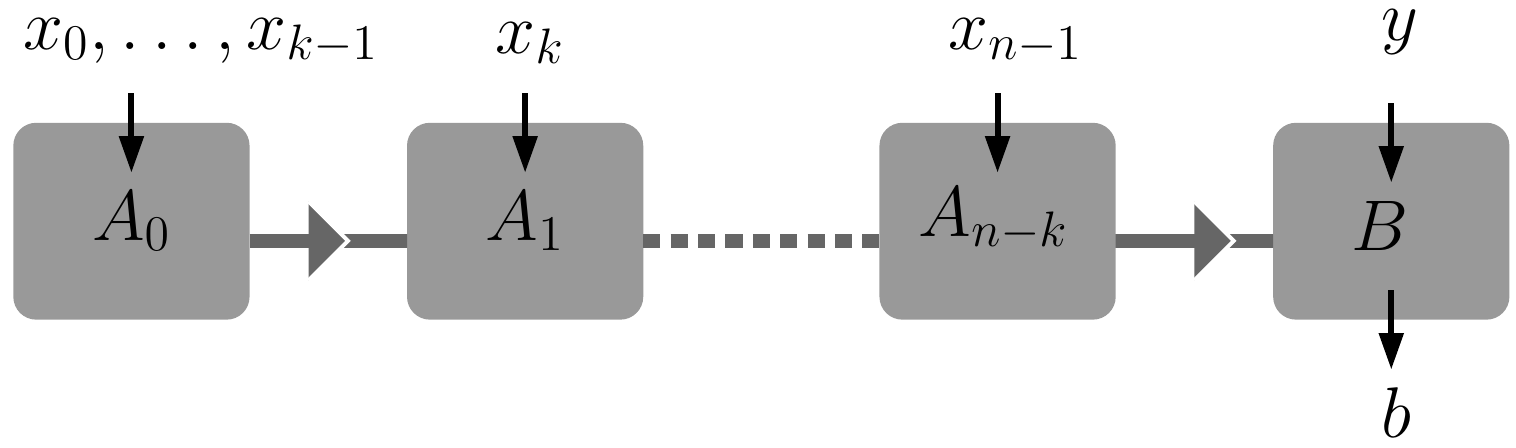}
\caption{ The general multiparty $(n,k)$-RAC scenario. Every channel carries no more than one bit or qubit of message in classical or quantum communication, respectively. }
    \label{figdef}
\end{figure}

In this work, we introduce the multiparty version of the RAC in which the $n$ bits of input are distributed among many parties that are arranged linearly. As shown in Fig.\ref{figdef}, the first party $A_0$ receives $k$ bits of input $x_0,\dots,x_{k-1} \in \{0,1\}^k$ and other parties $A_1,\dots,A_{n-k}$ receive one bit of input $x_k,\dots,x_{n-1}$ each, respectively. While the last party $B$ receives one number $y$ from the set $\{0,\dots,n-1\}$, and returns an binary output $b \in \{0,1\}$. The constraints are: (1) each party communicates only to the subsequent party, and (2) the channel capacity of every communication is no more than one. The aim is to recover partial information of the $n$ bits of input depending on $y$. Precisely, all the parties co-operate to optimize the average success probability, 
\begeq \label{P}
P = \frac{1}{n2^n} \sum_{x,y}  p(b=f(x,y)|x,y),
\endeq 
of returning a bi-variate function $f(x,y):= f(x_0,\dots,x_{n-1},y)$, where $x$ denotes the string of $n$ bits of input $x_0,\dots,x_{n-1}$, and $p(b|x,y)$ denotes the probability of returning $b$ given inputs $x,y$. 
We denote this task by $(n,k)$-RAC. Thus, $(n,n)$-RAC amounts to the standard two-party RAC. First, we consider the usual task of retrieving one of the $n$ bits distributed among $n-1$ parties, that is, $k=2$ and $f(x,y)=x_y$. We present two distinct EARAC protocols  applicable for any number of parties that involve sharing of two-qubit Bell states \cite{bell,chsh} and three-qubit Greenberger-Horne-Zeilinger (GHZ) states \cite{ghz}, respectively. Next, we outline a general approach to construct QRAC schemes in the multiparty scenario and provide few explicit examples of QRAC protocols. We also compare the success probability of the respective tasks in quantum protocols with their classical counterparts.

\section{Multiparty EARAC protocols}

The general EARAC protocols are constructed using a method, namely, \textit{concatenation} of known EARAC protocol in simplest scenarios \cite{earac,drac}.  We first formulate a class of concatenation scheme inspired by the concatenation protocol introduced in \cite{earac}. Subsequently, taking $(2,2)$-EARAC and $(3,2)$-EARAC \cite{earac,drac} as the primitives of our concatenation scheme, we demonstrate the generalized $(n,2)$-EARAC protocols.

\subsection{Concatenation of entanglement assisted protocol}

\begin{figure}[H]
\centering
\includegraphics[width=1\columnwidth]{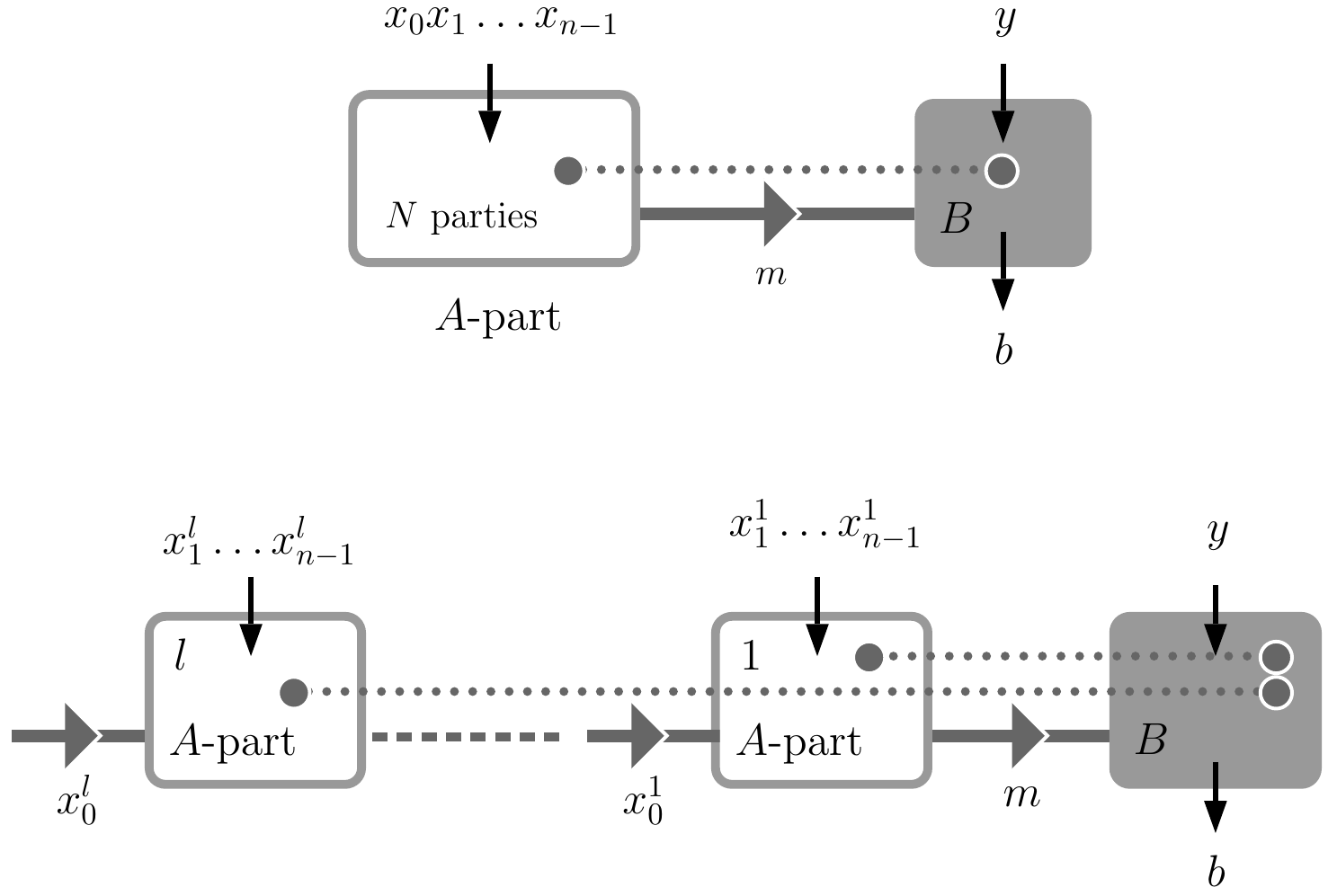}
\caption{An entanglement assisted quantum protocol $\mathcal{Q}(n,N,\epsilon)$ is shown. The $n$ bits of input are distributed to $N$ parties which we refer to as the $A$-part. An arbitrary $(N+1)$-partite entangled state, depicted by the dotted line, is shared between the $A$-part and the guessing party $B$. The concatenation scheme is presented below by taking $\mathcal{Q}(n,N,\epsilon)$ as the primitive. The first bit $x_0^l$ of $A$-part with level $l$ is the message from the succeeding $A$-part. }
\label{fig:c}
\end{figure}

Suppose we have an entanglement assisted quantum protocol $\mathcal{Q}(n,N,\epsilon)$ of the conventional RAC task of retrieving one of the $n$ bits, i.e., $f(x,y)=x_y$, with success probability $P=\frac{1}{2}(1+\epsilon)$. Note that $\epsilon$ is non-negative since the success probability of guessing one bit is at least $1/2$.  In general, the inputs are distributed among  $N$ number of parties arranged in an arbitrary manner, which we refer to as the $A$-part of the task. Only one party from the $A$-part is allowed to communicate one bit of message, say $m$, to the guessing party $B$ (see Fig. \ref{fig:c}). All the parties including the guessing party share arbitrary entangled states. Depending on the input and received messages from others, each party in $A$-part measures their respective subsystems and communicates the outcome.  The protocol $\mathcal{Q}(n,N,\epsilon)$ is such that, finally, depending on the input $y$, $B$ performs a binary outcome measurement on his shared quantum state resulting an outcome $c$ and his guess for the bit $x_y$ is $b = c\oplus m$. \\
Taking $\mathcal{Q}(n,N,\epsilon)$ as the primitive, a class of concatenation protocol can be formulated as follows. 
As shown in Fig. \ref{fig:c}, the concatenation is linear in the sense that the first bit of the input string comes from the preceding $A$-part as the message. Let $l$ denotes the number of communication channels that separates an $A$-part from the guessing party $B$. In other words, an $A$-part with level $l$ is connected to the guessing party $B$ via $(l-1)$ number of $\mathcal{Q}(n,N,\epsilon)$ protocols. The $(n-1)$bits of inputs received by that $A$-part is denoted by $x^l_1\dots x^l_{n-1}$. To guess one bit, say $x^l_i$, the guessing party completes $l$ number of $\mathcal{Q}(n,N,\epsilon)$ protocols. The first $(l-1)$ protocols are performed to obtain the messages $x^1_0,\dots, x^{l-1}_0$ that connects the relevant $A$-part to $B$, and subsequently, the last one is performed to obtain $x^l_i$. The guess is correct either all the  $\mathcal{Q}(n,N,\epsilon)$ protocols provide the correct outputs or the number of errors occurred in those $l$ protocols is even.  Thus, the success probability of guessing $x^l_i$ is given by,
\begeq \label{cp}
 \sum^{\lfloor \frac{l}{2}\rfloor}_{j=0} {l\choose{2j}} \left(\frac{1}{2}(1+\epsilon)\right)^{l-2j} \left(\frac{1}{2}(1+\epsilon)\right)^{2j} = \frac{1}{2}(1+\epsilon^{l}) .
 \endeq

Let us remark that the concatenation scheme proposed in \cite{earac} is restricted to two-party primitive RAC protocols, while the proposed one employs a primitive protocol of arbitrary number of parties. Furthermore, the above scheme is proposed in different scenario than the one considered in \cite{earac}.

\subsection{Concatenation protocol using Bell states}	
\begin{figure}[H]
\centering
\includegraphics[width=1\columnwidth]{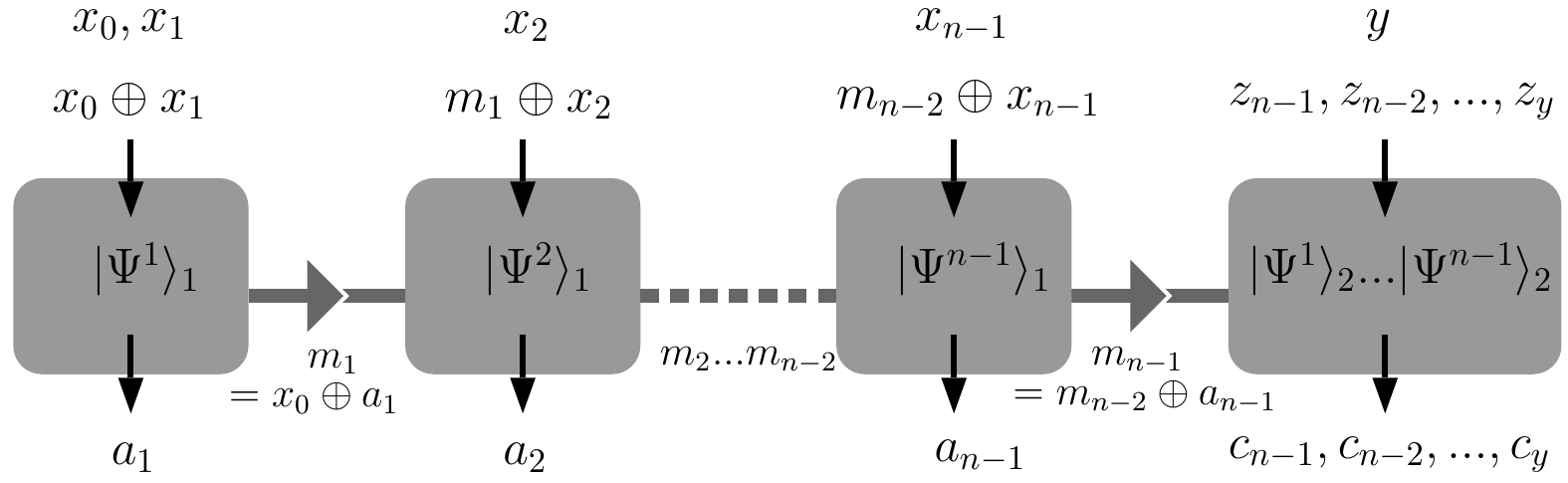}
\caption{ $\ket{\Psi^i}_j$ denotes the $j$-th particle of the $i$-th Bell state $\ket{\Psi}$. Each party $A_i$ shares one Bell state $\ket{\Psi^{i+1}}$ with the guessing party. The protocol is as follows: depending on $m_i\oplus x_{i+1}$ ($x_0\oplus x_1$ for $A_0$), $A_i$ measures one of the two dichotomic observables on $\ket{\Psi^{i+1}}_1$ which yields binary outcome $a_{i+1}\in \{0,1\}$. Then $A_i$ communicates a message $m_i\oplus a_{i+1}$ ($x_0\oplus a_1$ for $A_0$) to the subsequent party $A_{i+1}$.  The measurement settings for every $A_i$ is the same. The measurement bases are given by: $\{\psi(\beta_+,0),\psi(\beta_-,\pi)\}$ and $\{\psi(\beta_-,0),\psi(\beta_+,\pi)\}$ for input $0$ and $1$, respectively, where $\psi(\theta,\phi) = \cos(\theta)\ket{0}+\sin(\theta)e^{i\phi}\ket{1}$, $\beta_{\pm} = \cos^{-1}\left( \sqrt{\frac{\sqrt{2}\pm 1}{2\sqrt2}} \right),$ and the first vector corresponds to outcome 0. The protocol is realized sequentially which starts from $A_0$ and continues to $A_{n-1}$ who sends the message $m_{n-1}$ to $B$. For a given input $y$, $B$ measures binary observables on his share of $(n-y)$ Bell states. Here $z_i\in \{0,1\}$ denotes the measurement settings performed on the $i$-th pair of Bell state $\ket{\Psi^i}_2$, and $c_i\in\{0,1\}$ denotes the respective outcome. For $y\neq 0$, $B$ chooses $z_{n-1}=...=z_{y+1}=0, z_y=1$; and for $y=0$, $B$ chooses $z_i=0, \forall i\in \{0,\dots,n-1\}$. The measurement settings $z_i=0$ or 1 correspond to the bases $\{\ket{+},\ket{-}\}$ or $\{\ket{0},\ket{1}\}$. The final guess for $x_y$ is given by $b = \oplus^{n-1}_{i=y} c_i \oplus m_{n-1}$.   }
\label{fig:bell}
\end{figure}

Due to \cite{earac}, we know there exists a $\mathcal{Q}(2,1,2^{-\frac{1}{2}})$ EARAC protocol. This two party protocol involves sharing a two-qubit singlet state (Bell state),
\begeq 
|\Psi \rangle _{12} = \frac{1}{\sqrt{2}}(|01\rangle - |10\rangle)_{12} . \endeq 
We will employ $\mathcal{Q}(2,1,2^{-\frac{1}{2}})$ as the primitive of the concatenation scheme described in the above section. The $(n,2)$-RAC can be perceived as the concatenation scheme (Fig. \ref{fig:c}) where the $A$-part with level $l$ contains one party who receives one bit input $x_{n-l}$.
From \eqref{cp} we know that the success probability of guessing $x_{n-l}$ is $\frac{1}{2}(1+2^{-\frac{l}{2}})$, since $x_{n-l}$ is separated from the guessing party by $l$ number of communication channels. Consequently, we obtain the average success probability \eqref{P} as follows,
\begin{equation}
\begin{split}
& P_{Q} = \frac{1}{n} \left( \sum^{n-1}_{l=1}\frac{1}{2}(1+2^{-\frac{l}{2}}) + \frac{1}{2}(1+2^{-\frac{n-1}{2}}) \right)
\\& = \frac{1}{2} + \frac{1+\sqrt{2}-2^{-\frac{n-2}{2}}}{2n} .
\end{split}
\end{equation}
The schematic representation of the explicit EARAC protocol is shown in Fig. \ref{fig:bell}.

\subsection{Concatenation protocol using GHZ states}	

In \cite{drac}, an EARAC protocol $\mathcal{Q}(3,2,3^{-\frac{1}{2}})$ has been proposed using a three qubit GHZ state
\begeq
\ket{\Phi}_{123}=\frac{1}{\sqrt{2}} (|000\rangle + |111\rangle)_{123},
\endeq
shared between three parties. Here, the $A$-part contains two parties. Taking $\mathcal{Q}(3,2,3^{-\frac{1}{2}})$ protocol as the primitive,
the general $(n,2)$-EARAC protocol for any odd $n$ yields larger success probability than the previous protocol. In this case, the $(n,2)$-RAC can be understood by the concatenation scheme where the $A$-part with level $l$ contains two parties each of them receives one bit, $x_{n-l}$ and $x_{n-l-1}$. As $x_{n-l}$ and $x_{n-l-1}$ are the inputs on the $A$-part of level $l$, it follows from \eqref{cp} that the success probability of guessing any of these inputs is $\frac{1}{2}(1+3^{-\frac{l}{2}})$. Hence, the overall success probability in this scenario,
\begin{equation} \label{ghzP}
\begin{split}
& P_{Q} = \frac{1}{n} \left( \sum^{\frac{n-1}{2}}_{l=1}(1+3^{-\frac{l}{2}}) + \frac{1}{2}(1+3^{-\frac{n-1}{4}}) \right)
\\& = \frac{1}{2} + \frac{1+\sqrt{3}-3^{-\frac{n-3}{4}}}{2n} .
\end{split}
\end{equation} 
To explicate the efficacy of concatenation scheme, we consider another example of RAC task, shown in Fig. \ref{c2} (b). In this task, there are nine inputs each of which is in level two in terms of the concatenation of $\mathcal{Q}(3,2,3^{-\frac{1}{2}})$ protocol. Hence, the success probability of this task is $\frac{1}{2}(1+3^{-1}) = \frac{2}{3}$. The explicit protocols are described in Fig. \ref{c2}.

\subsection{Classical strategy}

\begin{figure}[H]
\centering
   \includegraphics[width=0.4\textwidth]{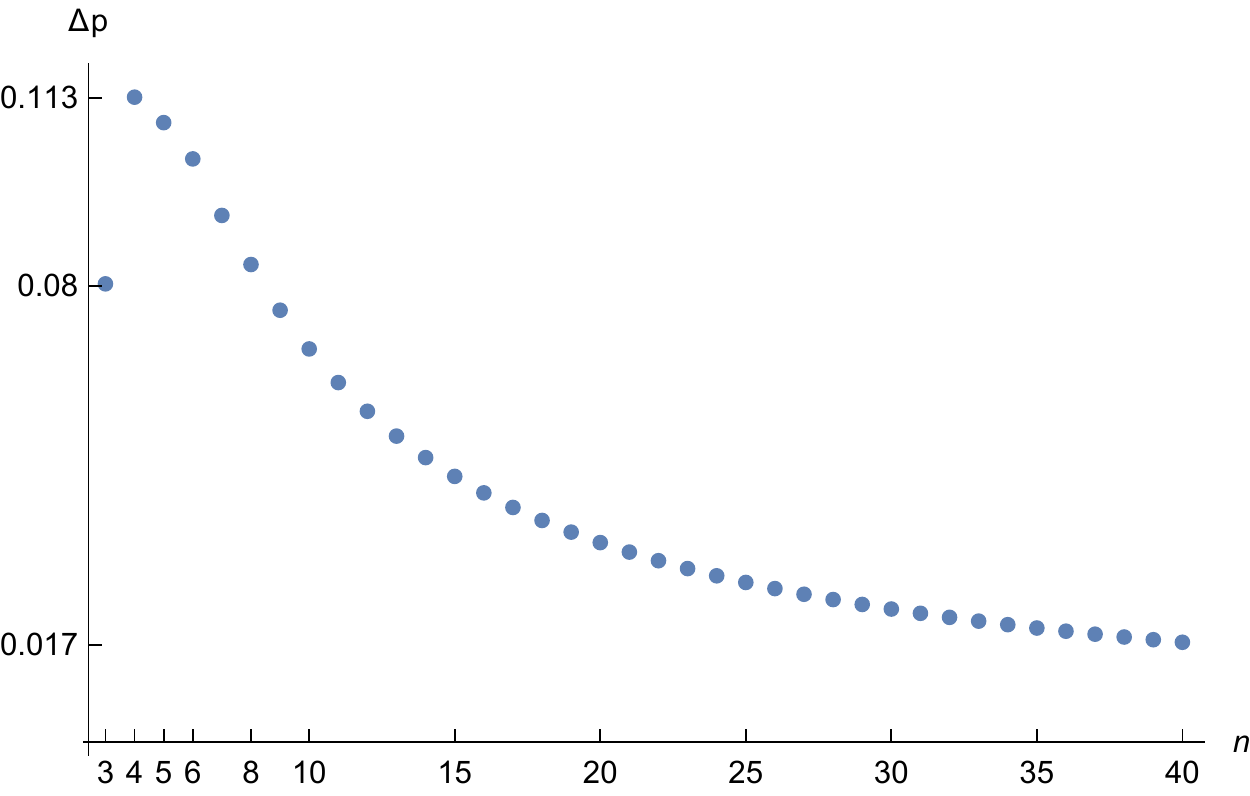}
    \caption{The difference between $P_{Q}$ \eqref{ghzP} and $P_C$ \eqref{eq:zigzag} is plotted. We observe the maximum difference 0.113 for $n=4$. }
    \label{fig:diplot}
\end{figure}

In this subsection, we discuss the classical counterpart of the multiparty RAC task. Since the average success probability \eqref{P} is a linear function of $p(z|x,y)$, it is  sufficient to consider only deterministic strategies to obtain the optimal success probability, say $P_C$, in classical communication. In $(n,2)$-RAC, each party $A_i$ receives eventually two bits of input - the message bit $m_i$ ($x_0$ for $A_0$) and the input bit $x_{i+1}$; and returns one bit of message $m_{i+1}$. Thus, any classical deterministic strategy for $A_i$ can be expressed by a function $\mathcal{E}: \{0,1\}\times \{0,1\} \rightarrow \{0,1\}$. There are $16$ different functions of this kind.
It can be shown that, without loss of generality, we can consider the strategy for the guessing party to be just returning the message, i.e., $b=m_{n-1}$ irrespective of the input $y$. For $n=3$, by considering all possible $\mathcal{E}$ for $A_0$ and $A_1$, we obtain the optimal success probability $\frac{17}{24}=0.708$. This value has been inappropriately stated in \cite{drac} (see Appendix \ref{app}). The corresponding strategies for $A_0$ and $A_1$ are $\mathcal{E}_1(x_0,x_1)=x_0 \cdot x_1 =m_1$ and $\mathcal{E}_2(m_1,x_2)=m_1\lor x_2 =m_2$, respectively, where $\lor$ denotes the `OR' operation. 
By exhausting all possible strategies,  it has been verified upto $n=8$ that the strategy yielding the optimal success probability follows the same pattern. Precisely, the optimal classical strategy is such that $m_i = m_{i-1} \cdot x_{i}$  for odd $i$ (here $m_0 \equiv x_0$), and   $m_i = m_{i-1} \lor x_{i}$ for even $i$.
The expression of the average success probability of $(n,2)$-RAC for such strategy,
\begin{equation}
\label{eq:zigzag}
\begin{split}
P_C &=\frac{1}{n2^n}\Bigg\{n^2+2 +\sum_{i=2}^{n-1}\bigg[\bigg(\binom{n-1}{i-1}+\sum_{k=2}^{i} \binom{n-2k+1}{i-k}\bigg)i\\
&+ \bigg(\binom{n}{i}-\binom{n-1}{i-1}-\sum_{k=2}^{i} \binom{n-2k+1}{i-k}\bigg)(n-i)\bigg]\Bigg\}.
\end{split}
\end{equation}
The above expression is obtained by dividing $n$-bit string into $n+1$ partitions such that each partition contains those strings that has $i$ number of 1's where $i$ runs from 0 to $n$, and subsequently computing the number of cases in which the output $b=1$ and 0 in each partition. A comparison between $P_C$ \eqref{eq:zigzag} and $P_Q$ \eqref{ghzP} pertaining to the EARAC protocol with GHZ state is shown in Fig. \ref{fig:diplot}.

\begin{widetext}

\begin{figure}[H]
\centering
\includegraphics[width=0.8\columnwidth]{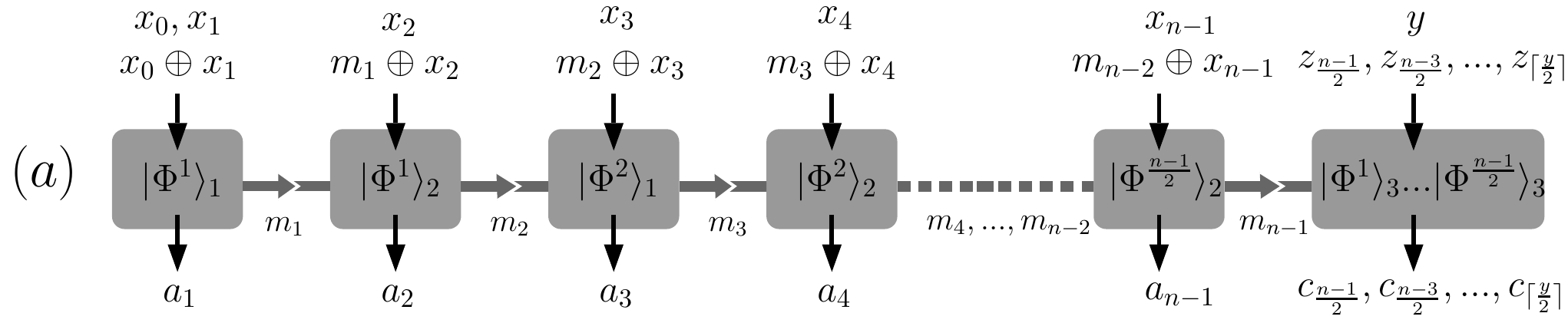} \\
\vspace{1cm}
\includegraphics[width=0.75\columnwidth]{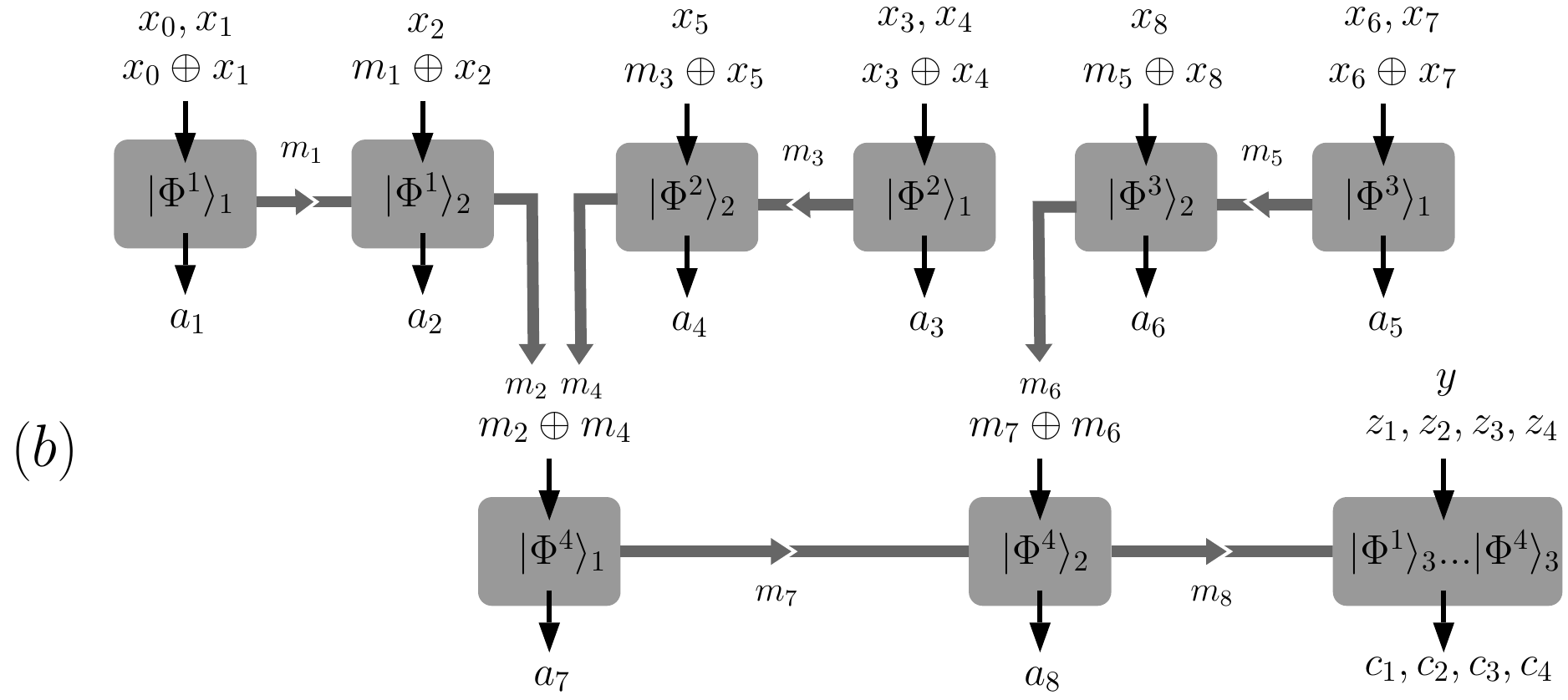}
\caption{$\ket{\Phi^i}_j$ denotes the $j$-th particle of the $i$-th GHZ state. The measurement settings on the respective subsystem are mentioned just above the upper arrow on each party; while the inputs are mentioned on the top of each party. In both the protocols, measurement settings on $\ket{\Phi^i}_1,\ket{\Phi^i}_2$ (for all $i$) are the same as in the primitive $\mathcal{Q}(3,2,3^{-\frac{1}{2}})$ EARAC \cite{earac}. On $\ket{\Phi^i}_1$, the measurement base is given by $\{\frac{1}{\sqrt{2}}(\ket{0}+e^{-i\phi}\ket{1}),\frac{1}{\sqrt{2}}(\ket{0}-e^{-i\phi}\ket{1})\}$ where $\phi=\frac{\pi}{4},\frac{3\pi}{4}$ for the input 0, 1 respectively. On $\ket{\Phi^i}_2$, the measurements basis is given by $\{\cos(\theta)\ket{0}+\sin{\theta}\ket{1},\sin(\theta)\ket{0}- \cos{\theta}\ket{1}\}$ where $\cos(\theta)=\sqrt{\frac{1}{2}+\frac{1}{2\sqrt{3}}},\sqrt{\frac{1}{2}-\frac{1}{2\sqrt{3}}}$ for the input 0,1 respectively. For the guessing party $B$, $z_i \in \{0,1,2\}$ denotes the measurement settings on $\ket{\Phi^i}_3$, and $c_i$ denotes the respective outcome. The settings $z_i=0,1,2$ correspond to the measurement bases $\{\ket{\circlearrowright}, \ket{\circlearrowleft}\}, \{\ket{+},\ket{-}\}$ and $\{\ket{0},\ket{1}\}$, respectively.\\ 
$(a)$ We consider $n$ is odd. Each party $A_{2i}$ (where $i=0,\dots,\frac{n-1}{2}$) shares one GHZ state $\ket{\Phi^{i+1}}$ with the subsequent party $A_{2i+1}$ and the guessing party $B$. The protocol is as follows: depending on $m_i\oplus x_{i+1}$ ($x_0\oplus x_1$ for $A_0$) $A_i$ measures one of the two dichotomic observables on the respective subsystem which yields binary outcome $a_{i+1}\in \{0,1\}$. As mentioned above, the measurement setting differs on the shared state $|\Phi\rangle_1$ or $|\Phi\rangle_2$. Then $A_i$ communicates a message $m_i\oplus a_{i+1}$ ($x_0\oplus a_1$ for $A_0$) to the subsequent party $A_{i+1}$. Upon receiving $y$ and $m_{n-1}$, the guessing party $B$ measures binary observables on $\lceil{\frac{n-y}{2}}\rceil$ different GHZ states. For $y\neq 0$, $B$ chooses $z_{\frac{n-1}{2}}=...=z_{\lceil{\frac{y}{2}}\rceil+1}=0,$ and $z_{\lceil{\frac{y}{2}}\rceil}=1$ (for odd $y$), or $ z_{\frac{y}{2}}=2$ (for even $y$). When $y=0$, $B$ chooses $\forall i \in \{0,\dots,\frac{n-1}{2}\},\ z_i=0 $.  The guess for $x_y$ is $b= \oplus^{\frac{n-1}{2}}_{i=\lceil{\frac{y}{2}}\rceil} c_i \oplus m_{n-1}$.\\
$(b)$ Here depending on the input $y$, measurements are performed on two GHZ states that connects $x_y$ to the guessing party. For example, when $y=3$, $B$ chooses $z_4=1,z_2=0$.
The guess for $x_y$ is $b=\oplus_i c_i \oplus m_8$, where $c_i$ are the outcomes of the relevant measurements.}

\label{c2}
\end{figure}

\clearpage

\end{widetext}

\section{Multiparty QRAC protocols }  

Given that there exists a $(n,n)$-QRAC protocol with success probability $P_Q$ where $f(x,y)=x_y$, we outline a general scheme to construct $(n,k)$-RAC task along with its QRAC protocol with the same success probability $P_Q$. As we aim to retain the same success probability of the $(n,n)$-QRAC protocol, all the $2^n$ states prepared by the sender should still be accessed by the guessing party $B$ in the $(n,k)$ scenario. The protocol is as follows. The first party $A_0$ prepares $2^k$ states according to his input $x_0,\dots,x_{k-1}$. Each party, $A_1,\dots,A_{n-k}$ applies a unitary, say $U_1,\dots,U_{n-k}$, respectively, on the received quantum system when the obtained input is 1, while they transit the received system to the subsequent party when the obtained input is 0. Depending on $x_k,\dots,x_{n-1}$ the total number of different unitaries, including the identity operation, act on the initial state before received by the guessing party is $2^{n-k}$. Thus, the required criteria are fulfilled if there exist unitaries $U_i$ for $i=1,\dots,n-k$ such that under those $2^{n-k}$ operations a subset of $2^k$ quantum states transform to all $2^n$ states which appeared in the $(n,n)$-QRAC protocol. However, the correspondence between the quantum state received by $B$ and the input string may differ from the standard $(n,n)$ scenario where $f(x,y)=x_y$. Suppose, the quantum state which is realized by $B$ for input $x = x_0\dots x_{n-1}$ encodes input $x' = x'_0\dots x'_{n-1}$ in the $(n,n)$-QRAC protocol. Then, we define the task as $f(x,y) = x'_y$ in the new $(n,k)$-RAC such that the same measurement settings on the guessing side yield the same success probability as before. Effectively, this construction allows us to  achieve the same average success probability as in the standard $(n,n)$-QRAC. Below we provide two examples of multiparty QRAC inspired from $(4,4)$ and $(6,6)$ QRAC protocols \cite{QRAC}.
We remark that, although, we use the same states and measurements of a $(n,n)$-QRAC protocol, the constructed multiparty protocol and the task are different than the original protocol. Moreover, not all $(n,n)$-QRAC protocol can be connected with the multiparty version.

\subsection{Construction of $(4,2)$-QRAC }

\begin{figure}[H]
\centering
   \includegraphics[width=0.35\textwidth]{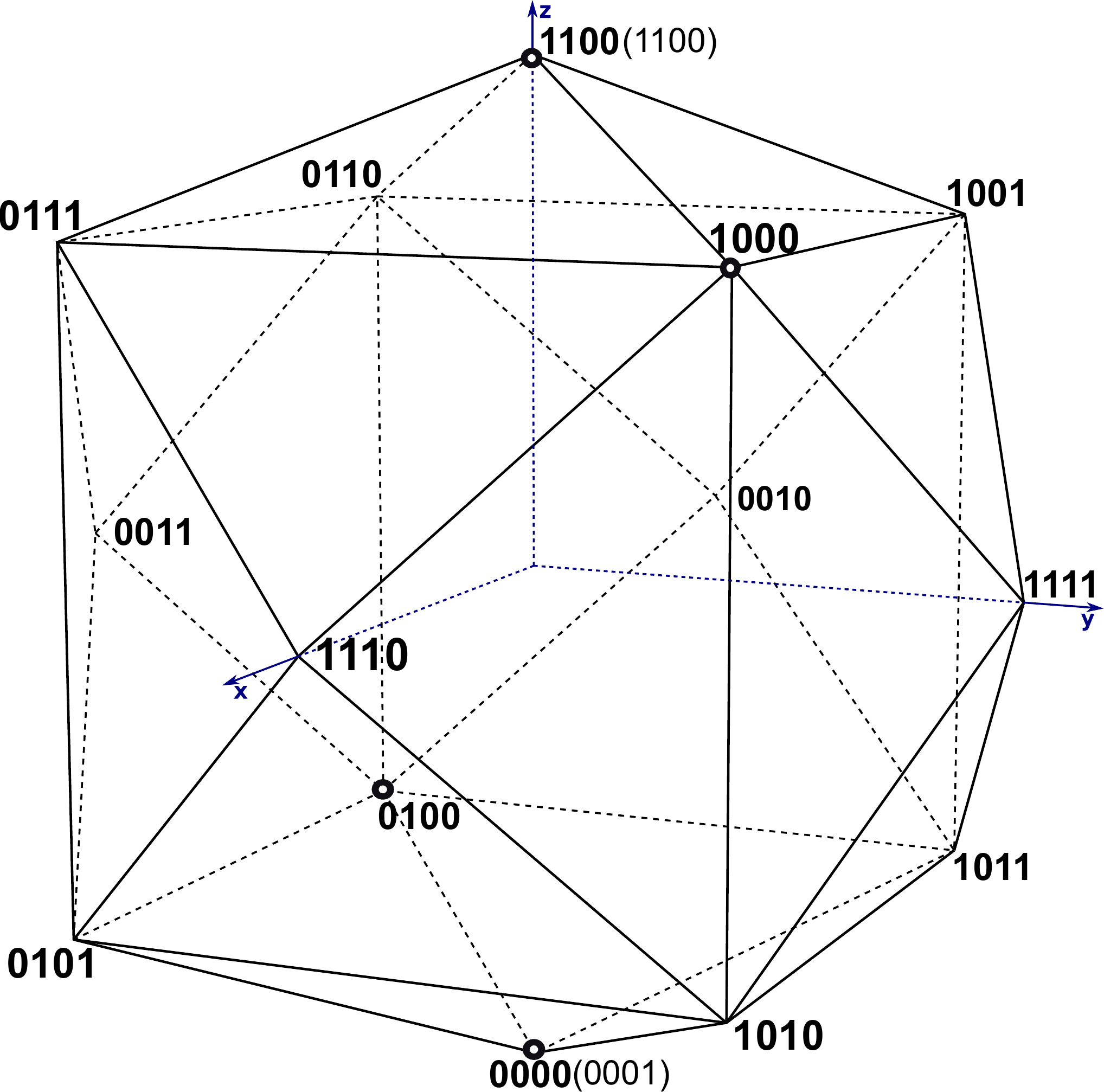}
    \caption{The encoding quantum states for different inputs in the $(4,2)$-QRAC protocol that correspond to the vertices of the \textit{Tetrakis Hexahedron} embedded in the Bloch sphere are presented. The four initial states prepared by $A_0$ are marked with the small circles. Unitaries used by $A_1,A_2$ are as follows: $U_1$ is $90^o$ counter clockwise rotation around $y$-axis and  $U_2$ is $90^o$ counter clockwise rotation around $z$-axis. We can verify that the set of four transformations $\{\mathbb{I},U_1,U_2,U_1U_2\}$ generate all the desired 16 states. Note that, we encode $14$ states in the vertices of the polyhedron and the remaining $2$ strings ($x = 1101$ and $x = 0001$) can be encoded anywhere without affecting the success probability. }
    \label{fig:THgame}
\end{figure}

We know a $(4,4)$-QRAC protocol with $P_Q= 0.733$ where the encoding states are the extreme points of \textit{ tetrakis hexahedron} (the convex hull of the cube and the octahedron) in the Bloch sphere \cite{QRAC}. Exploiting the nice symmetry possessed by these encoding states, we construct a $(4,2)$-QRAC protocol following the method described above. In the $(4,2)$ scenario, we choose two unitary transformations that together with the identity transformation construct the set of four transformations $\{\mathbb{I},U_1,U_2,U_1U_2\}$ which corresponds to $x_2x_3=00,01,10,11$, respectively. As shown in Fig. \ref{fig:THgame},  by choosing proper four vertices and unitaries $U_1,U_2$ we are able to arrive at all 16 states and therefore reconstruct the effective $(4,4)$-QRAC protocol presented in \cite{QRAC}. The rule of the new task is given in Table \ref{tab:tranformation}.  The detailed protocol is presented in Fig. \ref{fig:THgame}. One can verify by considering all possible deterministic classical strategies that $P_C=0.6562$ of the $(4,2)$-RAC task (given in Table \ref{tab:tranformation}), which is lower than $P_C = 0.6875$ in the standard scenario. Hence, the gap between the success probability of quantum and classical protocol  increases.


\begin{table}[H]
\centering
\resizebox{\columnwidth}{!}{%
\begin{tabular}{|c|c|c|c|}
\cline{1-1}
\multirow{1}*{List$1$ ($x$) $\rightarrow$ List$2$ ($x'$)}    \\ \cline{1-4}
$0000 \rightarrow 0011$  & $0010 \rightarrow 0101$ & $0001 \rightarrow 0000$ & $0011 \rightarrow 0110$ \\ \cline{1-4}
\multicolumn{1}{ |c| }{$0100 \rightarrow 0111$} & $0110 \rightarrow 0100$ & $0101 \rightarrow 0010$ & $0111 \rightarrow 1110$    \\ \cline{1-4}               
\multicolumn{1}{ |c| }{$1100 \rightarrow 1100$} & $1110 \rightarrow 1010$ & $1101 \rightarrow 1111$ & $1111 \rightarrow 1001$  \\ \cline{1-4}
\multicolumn{1}{ |c| }{$1000 \rightarrow 1000$} & $1010 \rightarrow 1011$ & $1001 \rightarrow 1101$ & $1011 \rightarrow 0001$ \\ \cline{1-4}
\end{tabular}
}
\caption{List$1$ represents the input string given to $A_i$, and List$2$ represents the function $f(x,y)$ $B$ wants to guess for input $y=0,1,2,3$. }
\label{tab:tranformation}
\end{table}


\subsection{Construction of $(6,3)$-QRAC } 
\begin{figure}[H]
\centering
\includegraphics[width=0.4\textwidth]{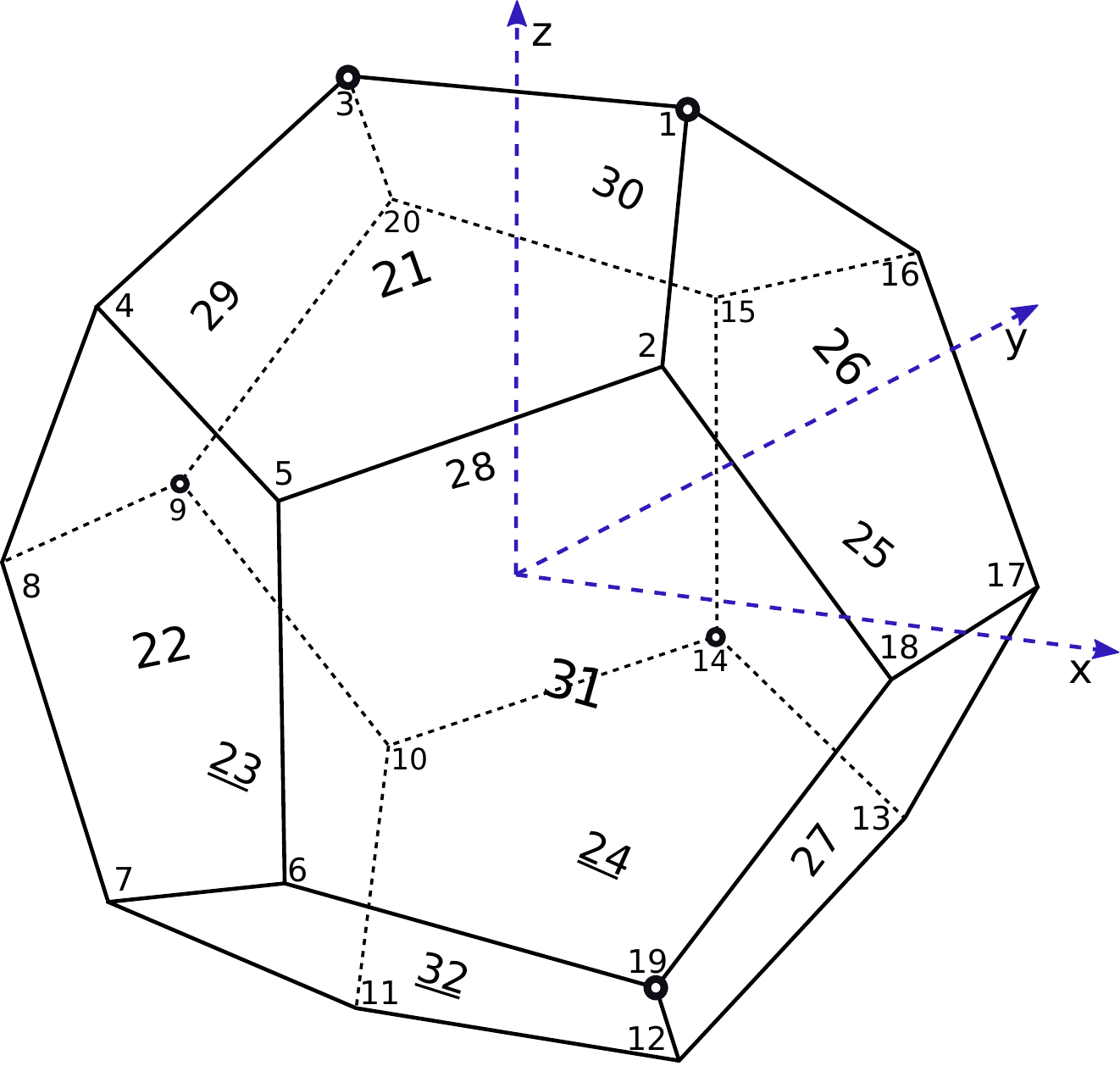}
\caption{For simplicity, we present here just the \textit{dodecahedron}. To construct \textit{pentakis dodecahedron} one has to build a pyramid on each of the pentagon faces of the \textit{dodecahedron}, and tops of those pyramids are the vertices (enumerated from 21 to 32) of the \textit{icosahedron}. The whole polyhedron is inscribed inside the Bloch sphere. The initial eight states prepared by $A_0$ that corresponds to input string $x_0x_1x_2$ are marked by the small circles (1,2,9,14,19) and the numbers with underline (23,24,32). We consider three unitaries for $A_1,A_2,A_3$ which forms the set of $8$ transformations $\{\mathbb{I}, U_1,U_2,U_3,U_1U_2,U_1U_3,U_2U_3,U_1U_2U_3 \}$, where $U_1,U_2,U_3$ are $180^o$ counter clockwise rotations around $x,y,z$ axis respectively. By applying these eight transformations on the initial states by $A_0$ we arrive at all the 32 vertices twice each, fulfilling the requirements of $(6,3)$-QRAC protocol.}
\label{fig:DO}
\end{figure}

The $(6,6)$-QRAC protocol proposed in \cite{QRAC} encodes $64$ qubit states into 32 vertices of \textit{pentakis dodecahedron}, which is geometrically the union of \textit{icosahedron} (12 vertices) and the \textit{dodecahedron} (20 vertices), embedded in the Bloch sphere. Thus, each vertex is associated with two inputs.  Measurement basis are defined along the six directions that are related to the vertices of the icosahedron that consist of six antipodal pairs. With this strategy we get the average success probability of $0.694$. We describe the QRAC scheme in Fig. \ref{fig:DO} that allows the guessing party $B$ to have access to all the 64 states. As stated before, accordingly, one can obtain the rule of the $(6,3)$-RAC task. The classical success probability is obtained to be $0.625$ while standard RAC scenario gives $0.65625$. 

\section{Conclusion}

In summary, we have introduced the multiparty version of random access codes and demonstrated several quantum protocols of it. Based on the proposed method of concatenation, we present entanglement assisted classical communication protocols applicable for arbitrary number of parties. We outline a general scheme for quantum multiparty RAC protocols along with a couple of examples. Interestingly, the advantage provided by quantum protocols over classical communication in terms of the average success probability is larger than the standard two-party scenario.

A further direction of research would be to generalize the presented multiparty QRAC protocols for larger numbers of $n$. The problem is related to the fact that whether there exists polyhedron embedded into a sphere with the necessary symmetries. Applicability of multiparty QRAC to quantum secure direct communication \cite{l1,l2,l3,l4,l5} in quantum network can be a very relevant future study. It will also be worthwhile to propose quantum key distribution, quantum randomness certification and self-testing of quantum devices in the multiparty scenario based on multiparty RAC.\\

\subsection*{Acknowledgement}
We are grateful to M. Paw\l owski for fruitful discussions. This work is supported by NCN grants 2016/23/N/ST2/02817,
2014/14/E/ST2/00020 and FNP grant First TEAM (Grant No. First TEAM/2017-4/31).

	\bibliographystyle{apsrev}
	\bibliography{bibliography}

\begin{thebibliography}{37}
\expandafter\ifx\csname natexlab\endcsname\relax\def\natexlab#1{#1}\fi
\expandafter\ifx\csname bibnamefont\endcsname\relax
  \def\bibnamefont#1{#1}\fi
\expandafter\ifx\csname bibfnamefont\endcsname\relax
  \def\bibfnamefont#1{#1}\fi
\expandafter\ifx\csname citenamefont\endcsname\relax
  \def\citenamefont#1{#1}\fi
\expandafter\ifx\csname url\endcsname\relax
  \def\url#1{\texttt{#1}}\fi
\expandafter\ifx\csname urlprefix\endcsname\relax\def\urlprefix{URL }\fi
\providecommand{\bibinfo}[2]{#2}
\providecommand{\eprint}[2][]{\url{#2}}

\bibitem[{\citenamefont{Holevo}(1973)}]{holevo}
\bibinfo{author}{\bibfnamefont{A.~S.} \bibnamefont{Holevo}},
  \bibinfo{journal}{Problemy Peredachi Informatsii}
  \textbf{\bibinfo{volume}{9}}, \bibinfo{pages}{3} (\bibinfo{year}{1973}).

\bibitem[{\citenamefont{Nielsen and Chuang}(2002)}]{book}
\bibinfo{author}{\bibfnamefont{M.~A.} \bibnamefont{Nielsen}} \bibnamefont{and}
  \bibinfo{author}{\bibfnamefont{I.}~\bibnamefont{Chuang}},
  \emph{\bibinfo{title}{Quantum computation and quantum information}}
  (\bibinfo{year}{2002}).

\bibitem[{\citenamefont{Ambainis et~al.}(1999)\citenamefont{Ambainis, Nayak,
  Ta-Shma, and Vazirani}}]{rac1}
\bibinfo{author}{\bibfnamefont{A.}~\bibnamefont{Ambainis}},
  \bibinfo{author}{\bibfnamefont{A.}~\bibnamefont{Nayak}},
  \bibinfo{author}{\bibfnamefont{A.}~\bibnamefont{Ta-Shma}}, \bibnamefont{and}
  \bibinfo{author}{\bibfnamefont{U.}~\bibnamefont{Vazirani}},
  \bibinfo{journal}{Proceedings of the 31st Annual ACM Symposium on Theory of
  Computing (STOC 99)} p. \bibinfo{pages}{376–383} (\bibinfo{year}{1999}).

\bibitem[{\citenamefont{Nayak}(1999)}]{rac2}
\bibinfo{author}{\bibfnamefont{A.}~\bibnamefont{Nayak}},
  \bibinfo{journal}{Proceedings of the 40th IEEE Symposium on Foundations of
  Computer Science (FOCS’99)} p. \bibinfo{pages}{369–376}
  (\bibinfo{year}{1999}).

\bibitem[{\citenamefont{Ambainis et~al.}(2002)\citenamefont{Ambainis, Nayak,
  Ta-Shma, and Vazirani}}]{rac3}
\bibinfo{author}{\bibfnamefont{A.}~\bibnamefont{Ambainis}},
  \bibinfo{author}{\bibfnamefont{A.}~\bibnamefont{Nayak}},
  \bibinfo{author}{\bibfnamefont{A.}~\bibnamefont{Ta-Shma}}, \bibnamefont{and}
  \bibinfo{author}{\bibfnamefont{U.}~\bibnamefont{Vazirani}},
  \bibinfo{journal}{Journal of the ACM} \textbf{\bibinfo{volume}{49}},
  \bibinfo{pages}{496–511} (\bibinfo{year}{2002}).

\bibitem[{\citenamefont{{Ambainis} et~al.}(2008)\citenamefont{{Ambainis},
  {Leung}, {Mancinska}, and {Ozols}}}]{QRAC}
\bibinfo{author}{\bibfnamefont{A.}~\bibnamefont{{Ambainis}}},
  \bibinfo{author}{\bibfnamefont{D.}~\bibnamefont{{Leung}}},
  \bibinfo{author}{\bibfnamefont{L.}~\bibnamefont{{Mancinska}}},
  \bibnamefont{and} \bibinfo{author}{\bibfnamefont{M.}~\bibnamefont{{Ozols}}},
  \bibinfo{journal}{ArXiv e-prints}  (\bibinfo{year}{2008}),
  \eprint{0810.2937}.

\bibitem[{\citenamefont{Paw\l{}owski and \ifmmode~\dot{Z}\else
  \.{Z}\fi{}ukowski}(2010)}]{earac}
\bibinfo{author}{\bibfnamefont{M.}~\bibnamefont{Paw\l{}owski}}
  \bibnamefont{and}
  \bibinfo{author}{\bibfnamefont{M.}~\bibnamefont{\ifmmode~\dot{Z}\else
  \.{Z}\fi{}ukowski}}, \bibinfo{journal}{Phys. Rev. A}
  \textbf{\bibinfo{volume}{81}}, \bibinfo{pages}{042326}
  (\bibinfo{year}{2010}).

\bibitem[{\citenamefont{Tavakoli et~al.}(2017)\citenamefont{Tavakoli,
  Paw\l{}owski, \ifmmode~\dot{Z}\else \.{Z}\fi{}ukowski, and
  Bourennane}}]{magic7}
\bibinfo{author}{\bibfnamefont{A.}~\bibnamefont{Tavakoli}},
  \bibinfo{author}{\bibfnamefont{M.}~\bibnamefont{Paw\l{}owski}},
  \bibinfo{author}{\bibfnamefont{M.}~\bibnamefont{\ifmmode~\dot{Z}\else
  \.{Z}\fi{}ukowski}}, \bibnamefont{and}
  \bibinfo{author}{\bibfnamefont{M.}~\bibnamefont{Bourennane}},
  \bibinfo{journal}{Phys. Rev. A} \textbf{\bibinfo{volume}{95}},
  \bibinfo{pages}{020302} (\bibinfo{year}{2017}).

\bibitem[{\citenamefont{Tavakoli et~al.}(2016)\citenamefont{Tavakoli, Marques,
  Paw\l{}owski, and Bourennane}}]{SvsS}
\bibinfo{author}{\bibfnamefont{A.}~\bibnamefont{Tavakoli}},
  \bibinfo{author}{\bibfnamefont{B.}~\bibnamefont{Marques}},
  \bibinfo{author}{\bibfnamefont{M.}~\bibnamefont{Paw\l{}owski}},
  \bibnamefont{and}
  \bibinfo{author}{\bibfnamefont{M.}~\bibnamefont{Bourennane}},
  \bibinfo{journal}{Phys. Rev. A} \textbf{\bibinfo{volume}{93}},
  \bibinfo{pages}{032336} (\bibinfo{year}{2016}).

\bibitem[{\citenamefont{Hameedi et~al.}(2017)\citenamefont{Hameedi, Saha,
  Mironowicz, Paw\l{}owski, and Bourennane}}]{drac}
\bibinfo{author}{\bibfnamefont{A.}~\bibnamefont{Hameedi}},
  \bibinfo{author}{\bibfnamefont{D.}~\bibnamefont{Saha}},
  \bibinfo{author}{\bibfnamefont{P.}~\bibnamefont{Mironowicz}},
  \bibinfo{author}{\bibfnamefont{M.}~\bibnamefont{Paw\l{}owski}},
  \bibnamefont{and}
  \bibinfo{author}{\bibfnamefont{M.}~\bibnamefont{Bourennane}},
  \bibinfo{journal}{Phys. Rev. A} \textbf{\bibinfo{volume}{95}},
  \bibinfo{pages}{052345} (\bibinfo{year}{2017}).

\bibitem[{\citenamefont{Paw\l{}owski and Brunner}(2011)}]{qkd}
\bibinfo{author}{\bibfnamefont{M.}~\bibnamefont{Paw\l{}owski}}
  \bibnamefont{and} \bibinfo{author}{\bibfnamefont{N.}~\bibnamefont{Brunner}},
  \bibinfo{journal}{Phys. Rev. A} \textbf{\bibinfo{volume}{84}},
  \bibinfo{pages}{010302} (\bibinfo{year}{2011}).

\bibitem[{\citenamefont{Bennett et~al.}(1991)\citenamefont{Bennett, Brassard,
  Cr{\'e}peau, and Skubiszewska}}]{qkd1}
\bibinfo{author}{\bibfnamefont{C.~H.} \bibnamefont{Bennett}},
  \bibinfo{author}{\bibfnamefont{G.}~\bibnamefont{Brassard}},
  \bibinfo{author}{\bibfnamefont{C.}~\bibnamefont{Cr{\'e}peau}},
  \bibnamefont{and} \bibinfo{author}{\bibfnamefont{M.-H.}
  \bibnamefont{Skubiszewska}}, in \emph{\bibinfo{booktitle}{Annual
  international cryptology conference}} (\bibinfo{organization}{Springer},
  \bibinfo{year}{1991}), pp. \bibinfo{pages}{351--366}.

\bibitem[{\citenamefont{Cr{\'e}peau}(1994)}]{qkd2}
\bibinfo{author}{\bibfnamefont{C.}~\bibnamefont{Cr{\'e}peau}},
  \bibinfo{journal}{Journal of Modern Optics} \textbf{\bibinfo{volume}{41}},
  \bibinfo{pages}{2445} (\bibinfo{year}{1994}).

\bibitem[{\citenamefont{Chaturvedi et~al.}(2018)\citenamefont{Chaturvedi, Ray,
  Veynar, and Paw{\l}owski}}]{sec}
\bibinfo{author}{\bibfnamefont{A.}~\bibnamefont{Chaturvedi}},
  \bibinfo{author}{\bibfnamefont{M.}~\bibnamefont{Ray}},
  \bibinfo{author}{\bibfnamefont{R.}~\bibnamefont{Veynar}}, \bibnamefont{and}
  \bibinfo{author}{\bibfnamefont{M.}~\bibnamefont{Paw{\l}owski}},
  \bibinfo{journal}{Quantum Information Processing}
  \textbf{\bibinfo{volume}{17}}, \bibinfo{pages}{131} (\bibinfo{year}{2018}).

\bibitem[{\citenamefont{Hameedi et~al.}(2015)\citenamefont{Hameedi, Marques,
  Mironowicz, Saha, Pawlowski, and Bourennane}}]{2s2r}
\bibinfo{author}{\bibfnamefont{A.}~\bibnamefont{Hameedi}},
  \bibinfo{author}{\bibfnamefont{B.}~\bibnamefont{Marques}},
  \bibinfo{author}{\bibfnamefont{P.}~\bibnamefont{Mironowicz}},
  \bibinfo{author}{\bibfnamefont{D.}~\bibnamefont{Saha}},
  \bibinfo{author}{\bibfnamefont{M.}~\bibnamefont{Pawlowski}},
  \bibnamefont{and}
  \bibinfo{author}{\bibfnamefont{M.}~\bibnamefont{Bourennane}},
  \bibinfo{journal}{arXiv preprint arXiv:1511.06179}  (\bibinfo{year}{2015}).

\bibitem[{\citenamefont{Li et~al.}(2011)\citenamefont{Li, Yin, Wu, Zou, Wang,
  Chen, Guo, and Han}}]{rc}
\bibinfo{author}{\bibfnamefont{H.-W.} \bibnamefont{Li}},
  \bibinfo{author}{\bibfnamefont{Z.-Q.} \bibnamefont{Yin}},
  \bibinfo{author}{\bibfnamefont{Y.-C.} \bibnamefont{Wu}},
  \bibinfo{author}{\bibfnamefont{X.-B.} \bibnamefont{Zou}},
  \bibinfo{author}{\bibfnamefont{S.}~\bibnamefont{Wang}},
  \bibinfo{author}{\bibfnamefont{W.}~\bibnamefont{Chen}},
  \bibinfo{author}{\bibfnamefont{G.-C.} \bibnamefont{Guo}}, \bibnamefont{and}
  \bibinfo{author}{\bibfnamefont{Z.-F.} \bibnamefont{Han}},
  \bibinfo{journal}{Phys. Rev. A} \textbf{\bibinfo{volume}{84}},
  \bibinfo{pages}{034301} (\bibinfo{year}{2011}).

\bibitem[{\citenamefont{Tavakoli et~al.}(2015)\citenamefont{Tavakoli, Hameedi,
  Marques, and Bourennane}}]{racd}
\bibinfo{author}{\bibfnamefont{A.}~\bibnamefont{Tavakoli}},
  \bibinfo{author}{\bibfnamefont{A.}~\bibnamefont{Hameedi}},
  \bibinfo{author}{\bibfnamefont{B.}~\bibnamefont{Marques}}, \bibnamefont{and}
  \bibinfo{author}{\bibfnamefont{M.}~\bibnamefont{Bourennane}},
  \bibinfo{journal}{Phys. Rev. Lett.} \textbf{\bibinfo{volume}{114}},
  \bibinfo{pages}{170502} (\bibinfo{year}{2015}).

\bibitem[{\citenamefont{Aguilar
  et~al.}(2018{\natexlab{a}})\citenamefont{Aguilar, Farkas, Mart\'{\i}nez,
  Alvarado, Cari\~ne, Xavier, Barra, Ca\~nas, Paw\l{}owski, and Lima}}]{EM}
\bibinfo{author}{\bibfnamefont{E.~A.} \bibnamefont{Aguilar}},
  \bibinfo{author}{\bibfnamefont{M.}~\bibnamefont{Farkas}},
  \bibinfo{author}{\bibfnamefont{D.}~\bibnamefont{Mart\'{\i}nez}},
  \bibinfo{author}{\bibfnamefont{M.}~\bibnamefont{Alvarado}},
  \bibinfo{author}{\bibfnamefont{J.}~\bibnamefont{Cari\~ne}},
  \bibinfo{author}{\bibfnamefont{G.~B.} \bibnamefont{Xavier}},
  \bibinfo{author}{\bibfnamefont{J.~F.} \bibnamefont{Barra}},
  \bibinfo{author}{\bibfnamefont{G.}~\bibnamefont{Ca\~nas}},
  \bibinfo{author}{\bibfnamefont{M.}~\bibnamefont{Paw\l{}owski}},
  \bibnamefont{and} \bibinfo{author}{\bibfnamefont{G.}~\bibnamefont{Lima}},
  \bibinfo{journal}{Phys. Rev. Lett.} \textbf{\bibinfo{volume}{120}},
  \bibinfo{pages}{230503} (\bibinfo{year}{2018}{\natexlab{a}}).

\bibitem[{\citenamefont{Bowles et~al.}(2015)\citenamefont{Bowles, Brunner, and
  Paw\l{}owski}}]{BBP}
\bibinfo{author}{\bibfnamefont{J.}~\bibnamefont{Bowles}},
  \bibinfo{author}{\bibfnamefont{N.}~\bibnamefont{Brunner}}, \bibnamefont{and}
  \bibinfo{author}{\bibfnamefont{M.}~\bibnamefont{Paw\l{}owski}},
  \bibinfo{journal}{Phys. Rev. A} \textbf{\bibinfo{volume}{92}},
  \bibinfo{pages}{022351} (\bibinfo{year}{2015}).

\bibitem[{\citenamefont{Czechlewski et~al.}(2018)\citenamefont{Czechlewski,
  Saha, Tavakoli, and Paw\l{}owski}}]{brac}
\bibinfo{author}{\bibfnamefont{M.}~\bibnamefont{Czechlewski}},
  \bibinfo{author}{\bibfnamefont{D.}~\bibnamefont{Saha}},
  \bibinfo{author}{\bibfnamefont{A.}~\bibnamefont{Tavakoli}}, \bibnamefont{and}
  \bibinfo{author}{\bibfnamefont{M.}~\bibnamefont{Paw\l{}owski}},
  \bibinfo{journal}{Phys. Rev. A} \textbf{\bibinfo{volume}{98}},
  \bibinfo{pages}{062305} (\bibinfo{year}{2018}).

\bibitem[{\citenamefont{Casaccino et~al.}(2008)\citenamefont{Casaccino,
  Galv\~ao, and Severini}}]{galvao}
\bibinfo{author}{\bibfnamefont{A.}~\bibnamefont{Casaccino}},
  \bibinfo{author}{\bibfnamefont{E.~F.} \bibnamefont{Galv\~ao}},
  \bibnamefont{and} \bibinfo{author}{\bibfnamefont{S.}~\bibnamefont{Severini}},
  \bibinfo{journal}{Phys. Rev. A} \textbf{\bibinfo{volume}{78}},
  \bibinfo{pages}{022310} (\bibinfo{year}{2008}).

\bibitem[{\citenamefont{Paw{\l}owski et~al.}(2009)\citenamefont{Paw{\l}owski,
  Paterek, Kaszlikowski, Scarani, Winter, and {\.Z}ukowski}}]{ic}
\bibinfo{author}{\bibfnamefont{M.}~\bibnamefont{Paw{\l}owski}},
  \bibinfo{author}{\bibfnamefont{T.}~\bibnamefont{Paterek}},
  \bibinfo{author}{\bibfnamefont{D.}~\bibnamefont{Kaszlikowski}},
  \bibinfo{author}{\bibfnamefont{V.}~\bibnamefont{Scarani}},
  \bibinfo{author}{\bibfnamefont{A.}~\bibnamefont{Winter}}, \bibnamefont{and}
  \bibinfo{author}{\bibfnamefont{M.}~\bibnamefont{{\.Z}ukowski}},
  \bibinfo{journal}{Nature} \textbf{\bibinfo{volume}{461}},
  \bibinfo{pages}{1101} (\bibinfo{year}{2009}).

\bibitem[{\citenamefont{Bobby and Paterek}(2014)}]{bobby}
\bibinfo{author}{\bibfnamefont{T.~K.~C.} \bibnamefont{Bobby}} \bibnamefont{and}
  \bibinfo{author}{\bibfnamefont{T.}~\bibnamefont{Paterek}},
  \bibinfo{journal}{New Journal of Physics} \textbf{\bibinfo{volume}{16}},
  \bibinfo{pages}{093063} (\bibinfo{year}{2014}).

\bibitem[{\citenamefont{Grudka et~al.}(2014)\citenamefont{Grudka, Horodecki,
  Horodecki, K\l{}obus, and Paw\l{}owski}}]{racbox}
\bibinfo{author}{\bibfnamefont{A.}~\bibnamefont{Grudka}},
  \bibinfo{author}{\bibfnamefont{K.}~\bibnamefont{Horodecki}},
  \bibinfo{author}{\bibfnamefont{M.}~\bibnamefont{Horodecki}},
  \bibinfo{author}{\bibfnamefont{W.}~\bibnamefont{K\l{}obus}},
  \bibnamefont{and}
  \bibinfo{author}{\bibfnamefont{M.}~\bibnamefont{Paw\l{}owski}},
  \bibinfo{journal}{Phys. Rev. Lett.} \textbf{\bibinfo{volume}{113}},
  \bibinfo{pages}{100401} (\bibinfo{year}{2014}).

\bibitem[{\citenamefont{Aguilar
  et~al.}(2018{\natexlab{b}})\citenamefont{Aguilar, Borka\l{}a, Mironowicz, and
  Paw\l{}owski}}]{prac}
\bibinfo{author}{\bibfnamefont{E.~A.} \bibnamefont{Aguilar}},
  \bibinfo{author}{\bibfnamefont{J.~J.} \bibnamefont{Borka\l{}a}},
  \bibinfo{author}{\bibfnamefont{P.}~\bibnamefont{Mironowicz}},
  \bibnamefont{and}
  \bibinfo{author}{\bibfnamefont{M.}~\bibnamefont{Paw\l{}owski}},
  \bibinfo{journal}{Phys. Rev. Lett.} \textbf{\bibinfo{volume}{121}},
  \bibinfo{pages}{050501} (\bibinfo{year}{2018}{\natexlab{b}}).

\bibitem[{\citenamefont{Saha et~al.}(2018)\citenamefont{Saha, Oszmaniec,
  Czekaj, Horodecki, and Horodecki}}]{opur}
\bibinfo{author}{\bibfnamefont{D.}~\bibnamefont{Saha}},
  \bibinfo{author}{\bibfnamefont{M.}~\bibnamefont{Oszmaniec}},
  \bibinfo{author}{\bibfnamefont{{\L}.}~\bibnamefont{Czekaj}},
  \bibinfo{author}{\bibfnamefont{M.}~\bibnamefont{Horodecki}},
  \bibnamefont{and}
  \bibinfo{author}{\bibfnamefont{R.}~\bibnamefont{Horodecki}},
  \bibinfo{journal}{arXiv preprint arXiv:1809.03475}  (\bibinfo{year}{2018}).

\bibitem[{\citenamefont{Farkas and Kaniewski}(2019)}]{st1}
\bibinfo{author}{\bibfnamefont{M.}~\bibnamefont{Farkas}} \bibnamefont{and}
  \bibinfo{author}{\bibfnamefont{J.}~\bibnamefont{Kaniewski}},
  \bibinfo{journal}{Phys. Rev. A} \textbf{\bibinfo{volume}{99}},
  \bibinfo{pages}{032316} (\bibinfo{year}{2019}).

\bibitem[{\citenamefont{Mironowicz and Paw\l{}owski}(2019)}]{st2}
\bibinfo{author}{\bibfnamefont{P.}~\bibnamefont{Mironowicz}} \bibnamefont{and}
  \bibinfo{author}{\bibfnamefont{M.}~\bibnamefont{Paw\l{}owski}},
  \bibinfo{journal}{Phys. Rev. A} \textbf{\bibinfo{volume}{100}},
  \bibinfo{pages}{030301} (\bibinfo{year}{2019}).

\bibitem[{\citenamefont{Miklin et~al.}(2019)\citenamefont{Miklin, Borka{\l}a,
  and Paw{\l}owski}}]{st3}
\bibinfo{author}{\bibfnamefont{N.}~\bibnamefont{Miklin}},
  \bibinfo{author}{\bibfnamefont{J.~J.} \bibnamefont{Borka{\l}a}},
  \bibnamefont{and}
  \bibinfo{author}{\bibfnamefont{M.}~\bibnamefont{Paw{\l}owski}},
  \bibinfo{journal}{arXiv preprint arXiv:1903.12533}  (\bibinfo{year}{2019}).

\bibitem[{\citenamefont{Bell}(1964)}]{bell}
\bibinfo{author}{\bibfnamefont{J.~S.} \bibnamefont{Bell}},
  \bibinfo{journal}{Physics} \textbf{\bibinfo{volume}{1}}, \bibinfo{pages}{195}
  (\bibinfo{year}{1964}).

\bibitem[{\citenamefont{Clauser et~al.}(1969)\citenamefont{Clauser, Horne,
  Shimony, and Holt}}]{chsh}
\bibinfo{author}{\bibfnamefont{J.~F.} \bibnamefont{Clauser}},
  \bibinfo{author}{\bibfnamefont{M.~A.} \bibnamefont{Horne}},
  \bibinfo{author}{\bibfnamefont{A.}~\bibnamefont{Shimony}}, \bibnamefont{and}
  \bibinfo{author}{\bibfnamefont{R.~A.} \bibnamefont{Holt}},
  \bibinfo{journal}{Phys. Rev. Lett.} \textbf{\bibinfo{volume}{23}},
  \bibinfo{pages}{880} (\bibinfo{year}{1969}).

\bibitem[{\citenamefont{Greenberger et~al.}(1989)\citenamefont{Greenberger,
  Horne, and Zeilinger}}]{ghz}
\bibinfo{author}{\bibfnamefont{D.~M.} \bibnamefont{Greenberger}},
  \bibinfo{author}{\bibfnamefont{M.~A.} \bibnamefont{Horne}}, \bibnamefont{and}
  \bibinfo{author}{\bibfnamefont{A.}~\bibnamefont{Zeilinger}},
  \emph{\bibinfo{title}{Bell's theorem, quantum theory, and conceptions of the
  universe}} (\bibinfo{year}{1989}).

\bibitem[{\citenamefont{Long and Liu}(2002)}]{l1}
\bibinfo{author}{\bibfnamefont{G.~L.} \bibnamefont{Long}} \bibnamefont{and}
  \bibinfo{author}{\bibfnamefont{X.~S.} \bibnamefont{Liu}},
  \bibinfo{journal}{Phys. Rev. A} \textbf{\bibinfo{volume}{65}},
  \bibinfo{pages}{032302} (\bibinfo{year}{2002}).

\bibitem[{\citenamefont{Deng et~al.}(2003)\citenamefont{Deng, Long, and
  Liu}}]{l2}
\bibinfo{author}{\bibfnamefont{F.-G.} \bibnamefont{Deng}},
  \bibinfo{author}{\bibfnamefont{G.~L.} \bibnamefont{Long}}, \bibnamefont{and}
  \bibinfo{author}{\bibfnamefont{X.-S.} \bibnamefont{Liu}},
  \bibinfo{journal}{Phys. Rev. A} \textbf{\bibinfo{volume}{68}},
  \bibinfo{pages}{042317} (\bibinfo{year}{2003}).

\bibitem[{\citenamefont{Deng and Long}(2004)}]{l3}
\bibinfo{author}{\bibfnamefont{F.-G.} \bibnamefont{Deng}} \bibnamefont{and}
  \bibinfo{author}{\bibfnamefont{G.~L.} \bibnamefont{Long}},
  \bibinfo{journal}{Phys. Rev. A} \textbf{\bibinfo{volume}{69}},
  \bibinfo{pages}{052319} (\bibinfo{year}{2004}).

\bibitem[{\citenamefont{Zhang et~al.}(2017)\citenamefont{Zhang, Ding, Sheng,
  Zhou, Shi, and Guo}}]{l4}
\bibinfo{author}{\bibfnamefont{W.}~\bibnamefont{Zhang}},
  \bibinfo{author}{\bibfnamefont{D.-S.} \bibnamefont{Ding}},
  \bibinfo{author}{\bibfnamefont{Y.-B.} \bibnamefont{Sheng}},
  \bibinfo{author}{\bibfnamefont{L.}~\bibnamefont{Zhou}},
  \bibinfo{author}{\bibfnamefont{B.-S.} \bibnamefont{Shi}}, \bibnamefont{and}
  \bibinfo{author}{\bibfnamefont{G.-C.} \bibnamefont{Guo}},
  \bibinfo{journal}{Phys. Rev. Lett.} \textbf{\bibinfo{volume}{118}},
  \bibinfo{pages}{220501} (\bibinfo{year}{2017}).

\bibitem[{\citenamefont{Qi et~al.}(2019)\citenamefont{Qi, Sun, Lin, Niu, Hao,
  Song, Huang, Gao, Yin, and Long}}]{l5}
\bibinfo{author}{\bibfnamefont{R.}~\bibnamefont{Qi}},
  \bibinfo{author}{\bibfnamefont{Z.}~\bibnamefont{Sun}},
  \bibinfo{author}{\bibfnamefont{Z.}~\bibnamefont{Lin}},
  \bibinfo{author}{\bibfnamefont{P.}~\bibnamefont{Niu}},
  \bibinfo{author}{\bibfnamefont{W.}~\bibnamefont{Hao}},
  \bibinfo{author}{\bibfnamefont{L.}~\bibnamefont{Song}},
  \bibinfo{author}{\bibfnamefont{Q.}~\bibnamefont{Huang}},
  \bibinfo{author}{\bibfnamefont{J.}~\bibnamefont{Gao}},
  \bibinfo{author}{\bibfnamefont{L.}~\bibnamefont{Yin}}, \bibnamefont{and}
  \bibinfo{author}{\bibfnamefont{G.-L.} \bibnamefont{Long}},
  \bibinfo{journal}{Light: Science \& Applications}
  \textbf{\bibinfo{volume}{8}}, \bibinfo{pages}{22} (\bibinfo{year}{2019}).

\end{thebibliography}

\begin{appendix}
	
\section{}
\label{app}

In Table I of \cite{drac}, a list of eight different tasks is presented in $(3,2)$ scenario. It has been stated that the success probability in classical communication is $\frac{2}{3}$ for all those eight cases. Here we correct that statement. The correct classical success probability obtained by considering all possible deterministic strategies for those eight tasks are given in table \ref{TabApB}.
	
\begin{table}[H]
\centering
\begin{tabular}{|l|l|}
\hline
$f(x,0),f(x,1),f(x,2)$  & $P_C$ \\ \hline
$x_1,x_2,x_3$  & $0.708$  \\ \hline
$x_0\oplus x_2,x_1\oplus x_2,x_2$      & $0.750$         \\ \hline
$x_0\oplus x_2(x_0\oplus x_1),$ &  \\                    
$x_1\oplus x_2(\overline{x_0\oplus x_1}),x_2$    &  $0.708$           \\ \hline
$x_0\oplus x_2(\overline{x_0\oplus x_1}),$ & \\
$x_1\oplus x_2(x_0\oplus x_1),x_2$   &  $0.708$           \\ \hline
$x_0 \oplus x_2,x_1,x_2$   & $0.708$              \\ \hline
$x_0,x_1,x_2\oplus x_0$    & $0.666$              \\ \hline
$x_0 \oplus x_2,x_1,x_0$    & $0.666$              \\ \hline
$x_0 \oplus x_2,x_1\oplus x_2,x_0$    & $0.666$              \\ \hline
\end{tabular}
\caption{List of optimal classical success probabilities for eight tasks in $(3,2)$-RAC scenario (Table I in \cite{drac}). The task is defined by $f(x,y)$ on the left column. }
\label{TabApB}
\end{table}
	
	\end{appendix}

\end{document}